\documentclass[apj]{emulateapj} \setkeys{Gin}{width=0.48\textwidth}
\usepackage{graphicx}

\newcommand{\hmpc}{\ifmmode{h^{-1}\,\hbox{Mpc}}\else{$h^{-1}$\thinspace Mpc}\fi}

\newcommand{\kms}{\ifmmode{\,\hbox{km\,s}^{-1}}\else {\rm\,km\,s$^{-1}$}\fi}
\newcommand{\msun}{{\rm\,M_\odot}}

\begin{document}
\title{Dynamical Simulations of the First Globular Clusters}
\shorttitle{First Globular Clusters}
\shortauthors{Carlberg}
\author{Raymond G. Carlberg}
\affil{Department of Astronomy \& Astrophysics, University of Toronto, Toronto, ON M5S 3H4, Canada} 
\email{carlberg@astro.utoronto.ca}

\begin{abstract}
A Milky Way-like halo is simulated with
tidally limited star clusters
inserted in the dark matter
halos present at high redshift. An n-body code augmented with
velocity relaxation in the star clusters evolves the system.
The stripped stars and remnant clusters that began in the lower mass sub-halos
have a distribution somewhat more extended than the dark matter
halo, with a mean galactic radius of about 60 kpc inside 150 kpc,
whereas the distribution of all stripped stars and clusters is more centrally concentrated than the dark matter.
The clusters from low mass sub-halos, those with a peak circular velocity of $ 12-18$ \kms,
also produce most of the population's thin stellar streams. Together these suggest a search strategy for 
extremely metal poor cluster stars and their remnants.
The dependence of the stellar population distribution on  sub-halo mass is not seen in simulations 
that start clusters at lower redshift.
The half mass radii of the clusters are set by the tidal fields of the initial sub-halo, ``dwarf galaxy'' location, which 
causes the average half mass radius to decrease with
increasing redshift of formation. Starting clusters at greater than
redshift 4 leads to cluster half mass radii
approximately as seen in the Milky Way. 
\end{abstract}

\keywords{Milky Way dynamics;  Milky Way dark matter halo}

\section{INTRODUCTION}

The first globular clusters can be defined as those that formed above redshift 6, approximately
the first billion years of the universe, the age over which reionization was largely completed \citep{PlanckXLVII:16}. 
Globular clusters may have played an important role 
in providing the ionizing photons for reionization \citep{Ricotti:02,BK:18}. Even if globular clusters 
are sub-dominant ionizing sources they are expected to be valuable tracers of star formation and structure 
at high redshift during their bright phase when O and B stars are present, being within the
observational  capabilities of the James Webb Space Telescope \citep{Carlberg:02,Renzini:17,BK:18}.
Confident identification of the remnants of the first clusters in current epoch galaxies, including the Milky Way,
will enable detailed studies of their chemical makeup at low metallicity,
which provides clues to the nature of even more metal poor stars, 
and to provide insights into the cluster's dynamical evolution and mass loss history. 

Globular clusters with metallicities below $\rm [Fe/H] \la -1$ have halo kinematics and
there is considerable evidence that these are predominantly accreted onto the Milky Way 
as part of the hierarchical buildup of the galaxy \citep{SZ:78,BS:06}.
Milky Way globular clusters show an age-metallicity relation \citep{VandenBerg:13}
for $\rm [Fe/H]\geq -1.7 $, but clusters with $\rm [Fe/H] \leq -1.7$ are
consistent with a common age that would place their formation in the $z>6$ regime.
In the Milky Way approximately 1/3 of the halo clusters have $\rm [Fe/H]\leq -1.7 $ and are within  one Giga-year
of the Hubble age, although the age dating uncertainties do not guarantee that they were formed at or above redshift 6.
Observational confirmation of the numbers of the  high redshift population is an important goal. Nevertheless,
the globular clusters that formed above redshift 6 likely are a population with significant numbers.

The oldest of the first globular clusters should  have very low metal abundances.
The metallicity distribution function of field halo stars is approximately in accord with
a simple closed box model \citep{Hartwick:76}, for $\rm [Fe/H] \ga -3.5$, 
below which the numbers begin to fall below the
relation \citep{DaCosta:19,Youakim:20}. 
On the other hand, globular clusters appear to exhibit a metallicity floor
at about $\rm [Fe/H] \simeq -2.5$ \citep{Beasley:19}. There is 
some theoretical support for  $\rm [Fe/H] \simeq -2.5$ being the minimum
to produce sufficient cooling to allow a globular cluster to form   \citep{Kruijssen:19}, but
it is dependent on abundances of individual elements at high redshift. A floor could be at a yet lower
metallicity \citep{Hartwick:18,Yoon:19}.  

The apparent metallicity floor for the globular clusters may indicate that extremely metal poor 
globular clusters did not form,
which would limit the usefulness of globular clusters as a tracer population of general star formation
at high redshift.
On the other hand, extremely metal poor globular clusters could have formed, then dissolved
in the tidal field of the galaxy and the progenitor systems that hosted them.  
Moreover, there remains the intriguing possibility that there are a  few of these rare objects 
waiting to be found. 
Encouragement comes from high resolution spectra of two stars in  inner halo Sylgr stream \citep{IMM:19}
which have essentially identical abundance
patterns at a metallicity of $\rm [Fe/H] = -2.9 $ \citep{RG:19}.
Since thin stellar streams generally arise from a globular cluster progenitor,
further confirmation of the Sylgr metallicity will push
the globular cluster metallicity floor down and increase the 
prospects for a successful hunt for extremely metal poor, $\rm [Fe/H]\la -3$, globular clusters and
additional extremely metal poor stellar streams.

The theoretical study of the formation of globular clusters 
has some basic physical considerations of sufficient gas cooling to allow stars to form
\citep{PD:68,Gunn:80}. The study of globular cluster formation in a cosmological context
 is coming  within the capabilities of high resolution numerical
simulations, although there is not yet good agreement on details.  
High redshift clusters likely form from the gas present in central regions of the relatively
low mass dark matter halos that are present at that time. There could be a 
single process, or, a 
two-stage globular cluster formation history \citep{KR:13}. 
Empirical study at low redshift finds that stars form in a wide range of group sizes and densities \citep{LL:03}.
The globular clusters that survive are likely to be the dense, massive cluster end of general star formation that
produces young clusters of a wide range of densities and masses,  \citep{BC:02,Leaman:13,Phipps:19}.
Tidal fields disperse most of the star clusters formed, leaving  those dense enough to survive in their 
tidal field environment \citep{FZ:01,Chandar:17}.
An alternate model emphasizes the creation of ``naked" of dark matter globular clusters in the collision of gas-rich dark matter
sub-halos which can produce the very high gas densities required for a globular cluster \citep{Trenti:15,LNJ:19,Madau:20}.
A key issue is the dominant physics that regulates the characteristic size of halo globular clusters. Tidal fields present during formation
and afterwards will always set an upper limit, but gas and radiation processes could produce smaller clusters. 

The evolution of globular clusters
within a cosmological distribution of dark matter 
is largely a dynamical problem.  The
dynamical phase begins after a complex and poorly understood phase 
of rapid star formation and vigorous stellar mass loss and gas expulsion in the first 100~Myr 
of their lifetime \citep{Calura:19,Howard:19}. 
The ubiquitous presence of distinct populations of chemically enriched second generation stars within globular clusters 
(summarized in \citet{Renzini:15} points to the complexity of the process but 
also suggests a cosmologically short timescale during which gas processes continued \citep{DAntona:16}.
The simulations here start when the clusters effectively become stellar dynamical objects, which
is sufficiently close in time to their star formation phase that they are near their site of formation.

One of the goals of this study is to provide some guidance on the likelihood of survival of 
very high redshift globular clusters and 
where the remnant clusters or tidal star streams are likely to be found. In a hierarchical assembly picture, the oldest, hence
most metal poor, stars in the
galactic halo should be centrally concentrated \citep{Tumlinson:10}, which is 
then likely true for globular clusters as a tracer population,
although globular clusters that orbit in the central few kilo-parsecs of the galaxy cannot survive long \citep{GO:99,GOT:14}.
Another goal is to follow the mass-radius relation of the clusters from their time of formation in the low mass dark halos present
at high redshift to their current epoch orbits in the dark matter halo of a Milky Way-like galaxy.

In this paper dynamically realistic globular clusters 
are started in a cosmologically realistic (cosmic web-like) distribution of dark matter
present at redshift 8. The dark matter and star clusters evolve under their 
gravitational forces to assemble into a Milky Way-like galactic halo, with most of the mass accretion before redshift one.
The analysis is supplemented with simulations started at redshift 4.6 and 3.2.
The simulations are useful to answer questions 
about the rate of mass loss from
the clusters subject to the strong tidal fields present, and, the galactic distribution of the clusters, their
half mass radii, and their tidally removed stars.
The next section briefly discusses the numerical methods.
Section \S3 discusses the time evolution of the simulations, \S4 the galactic distribution of clusters and stripped stars, 
comparing those from low and higher mass sub-halos. In \S5 the relationship
between the final cluster sizes and the starting conditions is explored.
The Discussion, \S6, proposes connections between these dynamical
results and the evolution and distribution of  metal poor globular clusters.

\begin{figure}
\begin{center}
\includegraphics[angle=0,scale=0.68]{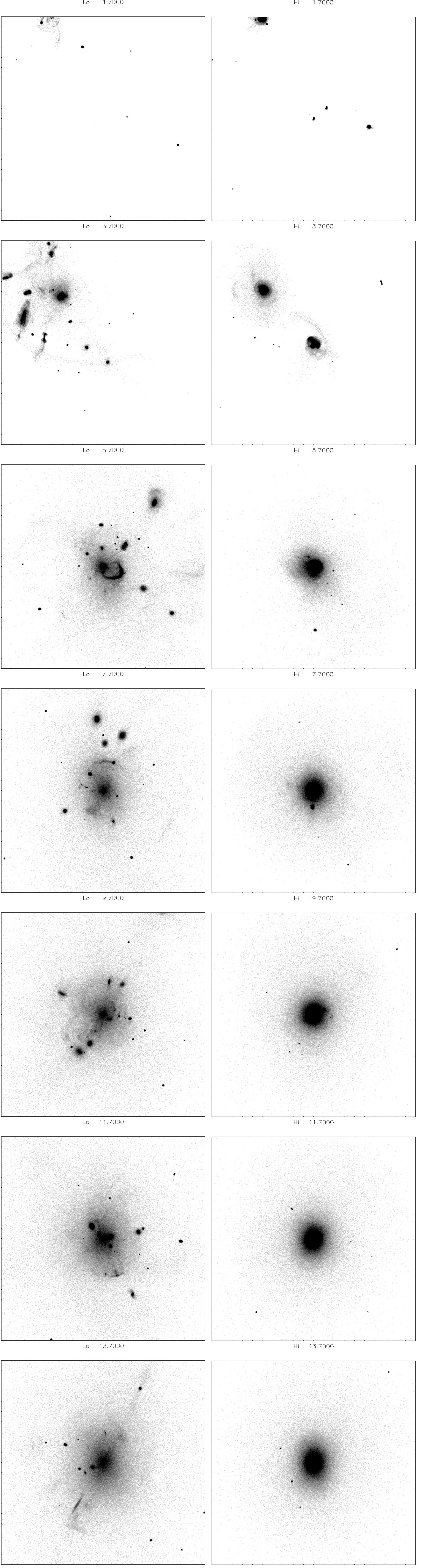}
\end{center}
\caption{
The evolving spatial distribution of globular cluster stars in a redshift 8 simulation.
Stars initially in clusters started in sub-halos below a virialized mass $2.5\times 10^8 \msun$ are
shown in the left panels and 
above   $25\times 10^8 \msun$, one decade in mass higher in the right panels, all 
 in a $\pm$100 kpc box. Times shown are 1.7, 3.7, 5.7, 7.7, 9.7, 11.7 and 13.7 (units of 1.022 Gyr) from the top.
Both sets of stars are from the same simulation.
}
\label{fig_lh100}
\end{figure}

\section{Cosmological Star Cluster Simulations}

The range of  about $10^5$ in dynamical length scale from star clusters, with a half mass
scale of 5-10 parsecs, to a galactic scale dark matter halo a virial radius of 200 kilo-parsecs is a numerical challenge.
The  stellar dynamics of stars clusters is fundamentally collisional \citep{Spitzer:87} in which gravitational 
interactions between
individual stars, binaries and other multiples allows some
stars to increase their orbital energy within the cluster at the expense of others. External
tidal fields heat and pull stars away from the clusters.
To tackle these problems a modified version of the standard cosmological n-body code, 
Gadget3 \citep{Gadget2} is employed without any hydrodynamics.
The dark
matter particles have a mass of $4\times 10^4\msun$ with a softening of 200 pc. 
Star particle clusters are added into the simulation. 
The star particles are made as close to stellar masses as practical, $5\msun$, with a softening of 2 pc.
in total there are about $10^8$ particles 
in the simulation, about 60\% of them dark matter. 

A Monte Carlo heating of the star particles in the clusters is added to Gadget3, in the
form of random velocity increments
of sizes determined by the half mass velocity relaxation rate \citep{Spitzer:87,BT:08}. 
The heating is calibrated 
using NBODY6 runs \citep{Aarseth:99, Carlberg:18, Carlberg:20}. Calibration runs of large N clusters
remain a challenge.  
 However, as cluster masses increase 
tidal heating dominates over internal heating, so the outcome of massive cluster evolution is expected not to 
depend sensitively on the star heating. 
Additional calibration and increasing the total number of particles to further refine and check the results will be valuable.

To create a Milky Way-like final system, the simulations
are conveniently initiated using the Via Lactea II halo catalogs \citep{VL2} which are reconstituted into particle halos
using the \citet{Hernquist:90} profile. The halos are traced back from the virialized halos at the final state 
of the simulation so need
to be boosted with extended halos that allows for the mass not associated with virialized sub-halos at redshift 8, 
for which the average extended
mass halo boosts the total mass a factor of 7-8.  The extended halos are about the same mass 
as the virialized core at redshift 4.6, and even less at redshift 3.2. The approach of reconstituting
halos offers some advantages and creates a realistic dark matter distribution, but is being pressed
to its limits at high redshift.

The star clusters are created using a King model equilibrium \citep{King:66} with a dimensionless
central potential, $W_0=7$. The outer radius of a cluster of mass $M_c$ is scaled to the simple tidal radius, 
$r_t = [M_c/(3 M(<r)]^{1/3} r$, where $M(<r)$ is the mass inside sub-halo at a cluster's initial orbital radius $r$. 
The clusters are usually set up to initially exactly fill this tidal radius. Tests show that tidally under-filling
initial clusters  quickly expand to the local tidal radius, whatever it is. For the chosen
value of $W_0$ the ratio of cluster half mass to tidal radius is then 0.22. 
Clusters are inserted into the sub-halo in a  disk-like distribution. The initial radii are drawn from 
an exponential disk distribution with a scale radius that is 20\% of the radius of the 
peak of the rotation curve for the sub-halo,
roughly what is expected based on the average angular momentum of the sub-halo. 
The number of clusters
is based on a ratio of cluster mass to virial mass of $10^{-4}$, which is about $1.3\times 10^{-5}$ allowing for 
the extended mass halo \citep{HHH:14}. There are typically one or two clusters per sub-halo. 
The clusters are started with a system velocity of the local circular velocity plus 5 \kms\ of random velocity.

\section{Evolution of the Clusters}

The simulations provide representative orbital histories of 
high redshift globular clusters that are accreted onto the Milky Way-like halo. 
Stars are pulled away from the clusters by the tidal fields in their
initial sub-halos, and later the main halo of the galaxy.  The orbits of 
the tidal stream stars are
accurately followed. The distribution of those tidal stars is one of the most important outcomes of
these simulations to guide searches for extremely metal poor clusters.

Figure~\ref{fig_lh100} shows the evolution of the spatial distribution of the clusters (unresolved in the plot) and stars that tidal fields
pull away from clusters. The left panel shows the stars from clusters originally in low mass sub-halos,
those with a virialized halo of less than $2.5\times 10^8 \msun$, total extended mass about $2\times 10^9 \msun$ or less.
The right panel shows the stars from clusters started in sub-halos at least ten times
more massive, $2.5\times 10^9 \msun$ and a total extended mass of about $2\times 10^{10} \msun$ and greater. 
The stars from clusters started in more massive sub-halos are much
more concentrated to the center of the dark halo in a nearly smooth density distribution, although one or two short 
streams appear near the center at late times. The clusters started in the low mass sub-halos produce a rich 
set of star streams, which are visible from the inner few kpc to 100 kpc. 
Because tidal fields pull stars away 
from a cluster near pericenter, at the inner turnaround point the stream width is comparable 
to the local tidal radius of the 
progenitor cluster, say 0.1 kpc. As the stream travels out to its outer turnaround radius
the width expands, often to a few kpc.

\begin{figure}
\begin{center}
\includegraphics[angle=-90,scale=0.7,trim=60 20 20 60, clip=true]{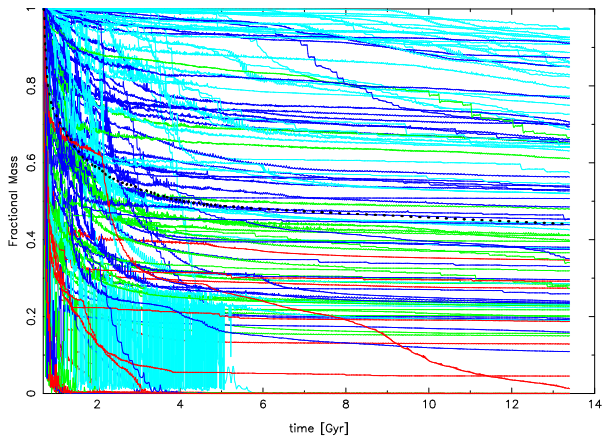}
\end{center}
\caption{The fractional mass remaining as a function of time
for clusters within 150 kpc of the main halo at the end of the
simulation. The colors scale gives the initial cluster mass, $M_c$, in $\log10{(M_c/\msun)}$ of the cluster,
ranging from turquoise (4.5-4.95), blue (4.95-5.4), green (5.4-5.85) and red (5.85-6.3). 
The black dotted line is the average. 
}
\label{fig_fmt}
\end{figure}

The time dependence of the fractional mass remaining in the clusters 
that are located within 150 kpc of the center of the halo at the end of the simulation is shown in Figure~\ref{fig_fmt}. 
On the average, the clusters started at high redshift lose a bit more than half of their mass of the
course of the simulation. Most of a cluster's mass is lost when it is in the strongest tidal fields at
the beginning of the simulation. 
The red lines are for the most massive clusters, which suffer the greatest mass loss.  
The problem of the massive clusters for high redshift of formation is an interesting one that 
will be considered in a future paper.

\begin{figure}
\begin{center}
\includegraphics[angle=-90,scale=0.7,trim=60 20 20 60, clip=true]{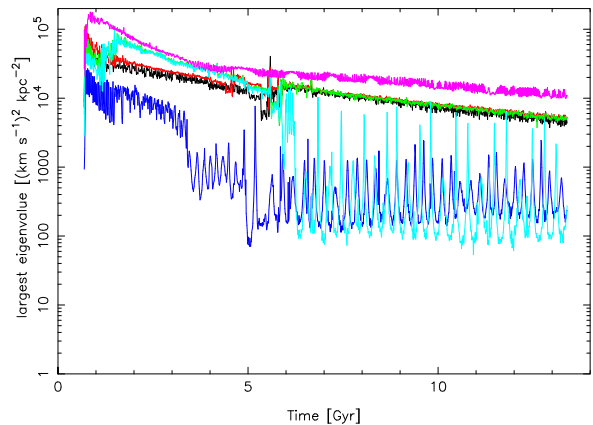} 
\includegraphics[angle=-90,scale=0.7,trim=60 20 20 60, clip=true]{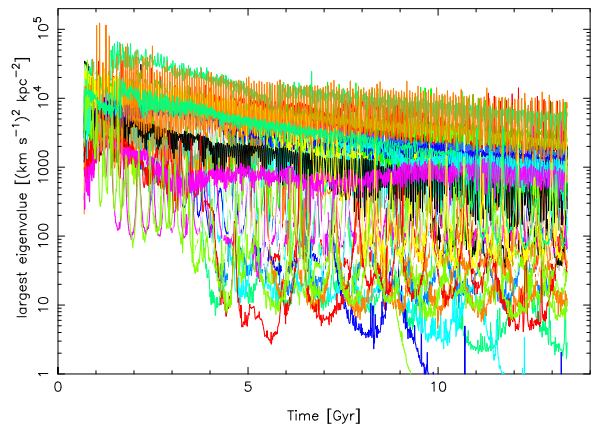} 
\end{center}
\caption{The largest eigenvalue of the tidal matrix  as a function of time for
the clusters in the high mass sub-halos (top panel)
and the low mass sub-halos (bottom panel).
}
\label{fig_tides}
\end{figure}

The tidal tensor is calculated as the second derivatives of the potential with respect to the coordinates, using
an offset from the cluster center of 50 pc. The  largest absolute eigenvalue of the tidal matrix is the plotted quantity.
Ongoing merging of the sub-halos into the growing
main halo causes the mean tidal fields to drop with time, Figure~\ref{fig_tides}. The tidal 
field in a dark  halo with circular velocity $V_c$ scales with $V_c^2/r^2$ at radius $r$. The star clusters
begin in dark halos with circular velocities of some 20 \kms\ which merge to create a galactic halo
with a circular velocity of 240 \kms. The decreasing tidal fields reflect the increase in orbital radius as a result of halo buildup.

The tidal field increases with distance from the center of a star cluster, driving an approximately
linear increase in impulsive velocity change in the outskirts of the cluster, or a quadratic increase in kinetic energy
with size. Although the clusters are set up to have about the same mean local density, $r\propto M^{0.33}$
the variation in location within a  sub-halo and that massive
clusters are more likely to be found in more massive sub-halos leads to an outcome where the population
of clusters has a size mass relation, 
$r\propto M^{0.6}$. High mass clusters are then  relatively more affected by tidal heating.

\begin{figure}
\begin{center}
\includegraphics[angle=-90,scale=0.7,trim=60 20 20 60, clip=true]{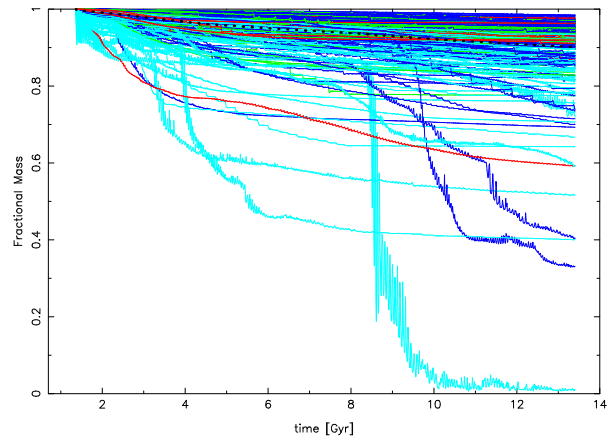}
\end{center}
\caption{The fractional mass remaining in the clusters 
as a function of time for a redshift 4.56 start. The colors are
the same as in Figure~\ref{fig_fmt}.
}
\label{fig_fmt358}
\end{figure}

The rate of mass loss drops very quickly with  time in Figure~\ref{fig_fmt}. Mass loss is largely the result of tidal heating,
which drop quickly as the substructure merges together to reduce the average density, hence tidal field
strength leads to a fairly rapid decline in the mass loss rate. 
A simulation started at redshift 4.56 with the same setup principles has
a much lower mass loss rate as shown in Figure~\ref{fig_fmt358}. The small dense halos
present in the redshift 8 start are a challenge to the numerical resolution, so there is also likely to be
some two-body relaxation between the heavy dark matter particles and the light star particles which
requires larger N simulations.

It is observationally well established that globular clusters lose mass to tidal fields, 
perhaps as best illustrated with the Pal~5 cluster \citep{GD:06} although there is an ever growing list of globular
clusters with clearly detected tidal tails.  
The chemically unique second generation stars are used to estimate
the fraction of the stellar halo that originated within globular clusters \citep{Martell:11} 
with current estimates being about 10\% \citep{Koch:19}. Their results
imply that clusters as a population have lost some 50+\% of their stellar mass 
and about half of the initially present clusters have dissolved.
The empirical estimates contain a number of parameters with uncertain values,
however, on the whole the substantial presence of stars originating in globular clusters
favors a redshift higher than 5 to give the strong tides required for the observationally
inferred halo fraction of globular cluster stars.

\section{The Distribution of Cluster Remnants}

\begin{figure}
\begin{center}
\includegraphics[angle=-90,scale=0.7,trim=60 20 20 60, clip=true]{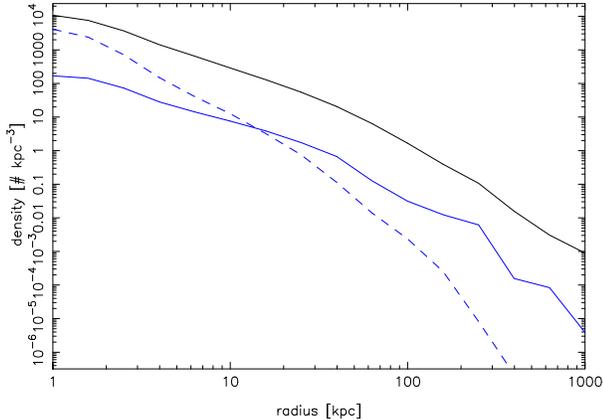}
\end{center}
\caption{
The radial density distribution of stars lost from clusters in
 sub-halos with mass below  $0.025 \times 10^{10}\msun$ (blue solid line) and the higher sub-halos 
mass above  $0.25 \times 10^{10}\msun$ (blue dashed line) at the
end of the simulation. 
The dark matter density profile is shown as the black solid line.
}
\label{fig_denz8}
\end{figure}

The number of unbound cluster stars per unit volume as a function of  radial distance in the halo 
 is shown in Figure~\ref{fig_denz8}. 
The lines show the densities of the stars from clusters in the lowest  mass sub-halos, 
those below $0.025 \times 10^{10}\msun$
(solid line)
and the highest mass sub-halos, those above $0.25 \times 10^{10}\msun$.

\begin{figure}
\begin{center}
\includegraphics[angle=-90,scale=0.7,trim=60 20 20 60, clip=true]{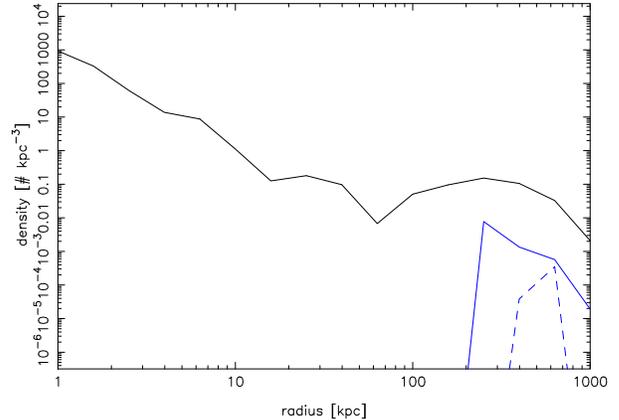}
\end{center}
\caption{
The radial density distribution of the stars lost from clusters in
low  (solid blue) and high (dashed blue) mass sub-halos near 
the beginning of the simulation, 0.73 Gyrs after the start. 
There is no initial difference in density distribution of the star
clusters with sub-halo mass.
}
\label{fig_den75}
\end{figure}

The much more concentrated density distribution that develops for the star clusters that begin in the more massive
sub-halos, see Figure~\ref{fig_lh100}, is not present in the early time distribution, although
the clusters are stronger tidal fields on the average, see Figure~\ref{fig_tides}. Figure~\ref{fig_den75} shows
the radial distribution of the same clusters as in Figure~\ref{fig_denz8}, but at time 1.45, the model
universe has an age of 1.41 Gyr and the simulation
is 0.73 Gyr old. The differences in radial distribution of the two sub-populations that develop over the course
of the simulation
is likely due to dynamical friction during infall being a larger effect for high mass sub-halo infall than for low mass sub-halos.

\begin{figure}
\begin{center}
\includegraphics[angle=-90,scale=0.7,trim=60 20 20 60, clip=true]{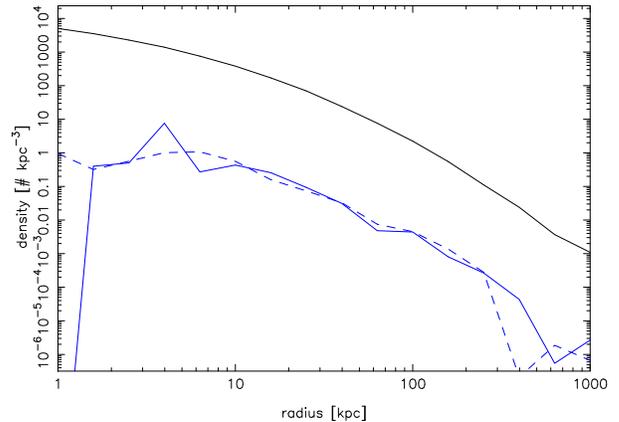} 
\end{center}
\caption{
Similar to Figure~\ref{fig_denz8} but for a redshift 4.56 start.
The radial density distribution of stars removed from clusters in
 sub-halos with mass below  $0.035 \times 10^{10}\msun$ (blue solid line) and the higher sub-halos 
mass above  $0.35 \times 10^{10}\msun$ (blue dashed line) at the
end of the simulation. 
The dark matter density profile is shown as the black solid line.
}
\label{fig_denz4}
\end{figure}

The segregation of clusters formed in low and high mass sub-halos is not prominent for
simulations started at
lower redshifts. Density profiles of dark matter, stars lost from clusters in sub-halos below $0.035\times 10^{10}\msun$
and stars from clusters in sub-halos with mass above $0.35\times 10^{10}\msun$
are shown in Figure~\ref{fig_denz4} for a simulation started at redshift 4.56.
Although the mass ranges are comparable to the higher redshift start, the density
profiles of the clusters in different mass sub-halos do not show a dependence on sub-halo mass. The
overall density distribution of this lower redshift infall population
is more extended than the dark matter. A redshift 3 start has an even
less centrally concentrated distribution of remnant clusters and stellar debris.

\begin{figure}
\begin{center}
\includegraphics[angle=-90,scale=0.7,trim=60 20 20 60, clip=true]{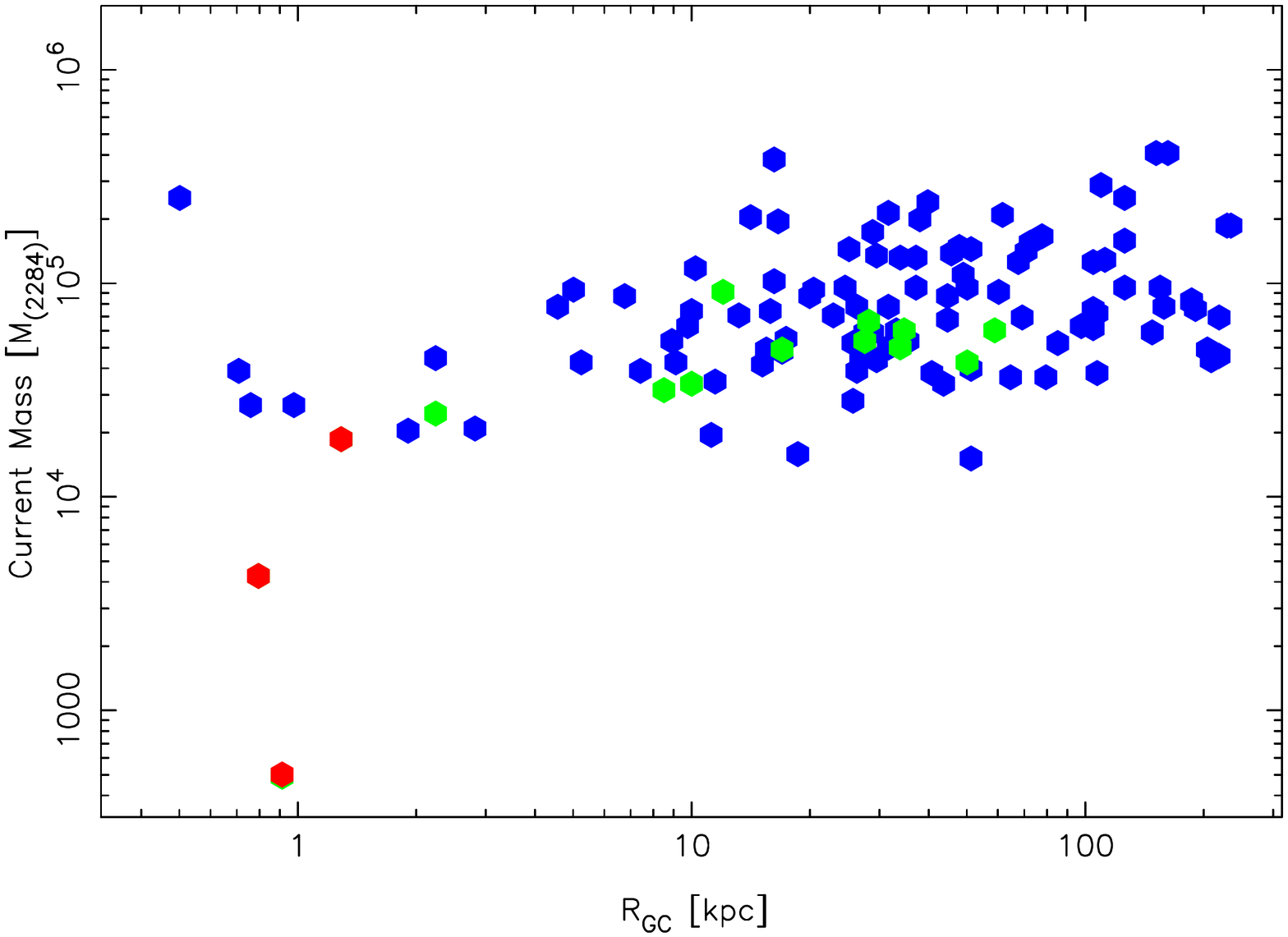} 
\put(-100,-90){\includegraphics[angle=-90,scale=0.25,trim=60 20 20 60, clip=true]{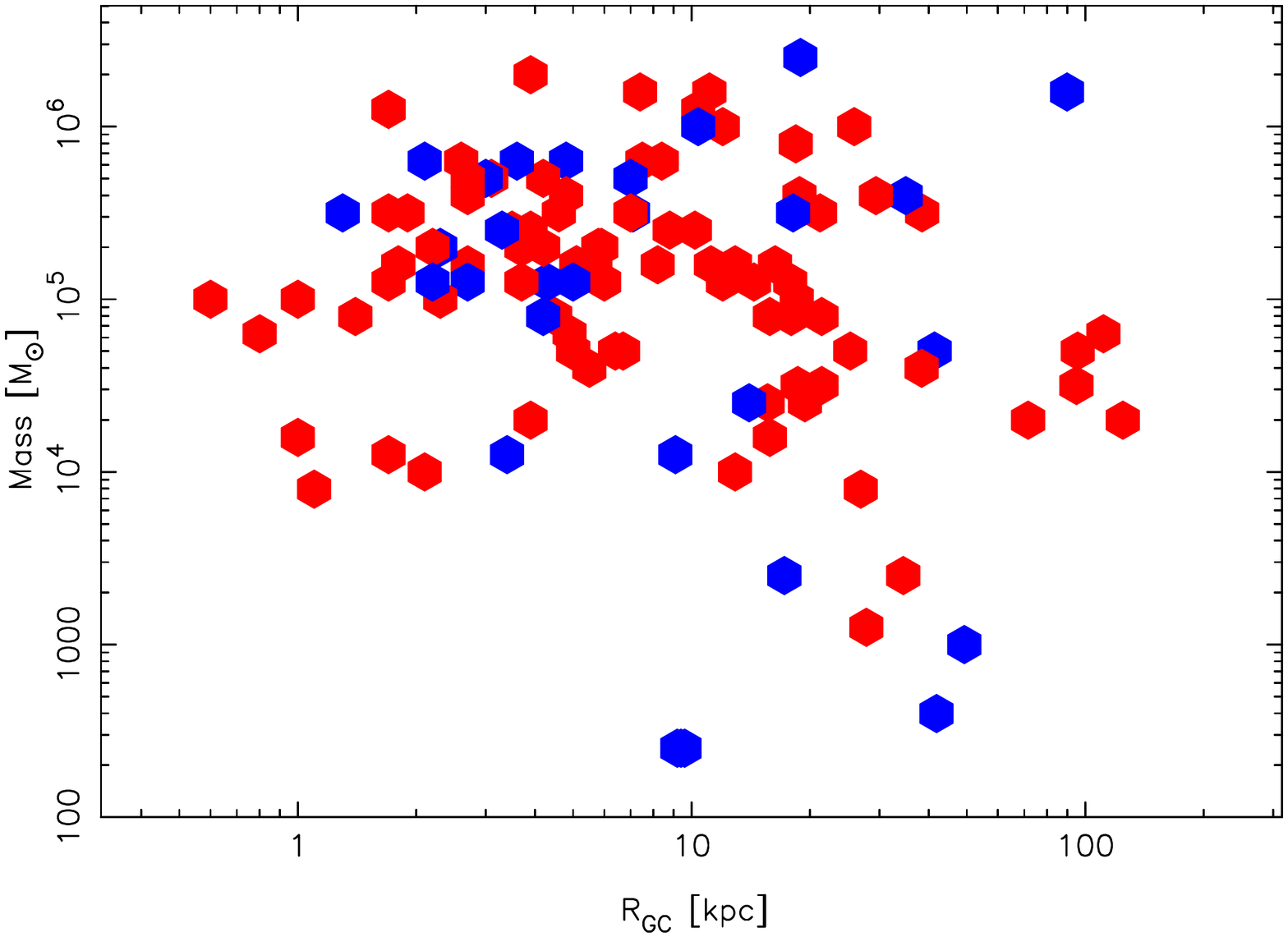}} 
\end{center}
\caption{
Current radial locations of clusters started in sub-halos of mass less than $2.5\times 10^8\msun$ (blue)
greater than $10\times 10^8\msun$ (green) and greater than $25\times 10^8\msun$ (red). 
Two of the five most massive clusters have completely evaporated, and there are four other clusters
inside 10 kpc that have lost so much mass that they  fall below the plotted mass range.
The inset shows the distribution of clusters in the \citet{Harris:96} catalog, assuming 
a mass-to-light of 2.
The colors of the inset indicate the metallicity range
$\rm -1 \geq [Fe/H]> -2$ (red)
$\rm [Fe/H]\leq -2$ (blue). 
}
\label{fig_rm}
\end{figure}

The endpoint cluster masses with their distances from the center in the
galactic halo are shown in Figure~\ref{fig_rm}. The numbers of clusters 
are based on starting with a ratio of cluster mass to halo mass of $1.2\times 10^{-5}$, which
may be appropriate for the entire halo cluster population, of which redshift 8 clusters would be some
currently unknown fraction.  
The red points in Figure~\ref{fig_rm} are the 3 surviving clusters that began in the six most massive sub-halos. They
are at such small galactic radii that they would certainly be merged into the baryonic disk and bulge.
The blue dots in Figure~\ref{fig_rm} are the clusters started in the 58 lowest mass sub-halos that were seeded with globular clusters. 
These clusters lose 
significant mass when within 20 kpc of the center, but are largely a more distant population. The
green dots are for a cluster population started in the 31 heaviest halos and are an intermediate
population of lower mass and smaller galactic radii.  The blue dots in the Figure~\ref{fig_rm} inset are the 
$\rm [Fe/H] <-2$ clusters in the \citet{Harris:96} catalog and appear to have a distribution
closest to the clusters started in the low mass sub-halos. The simulated clusters do not
have final time clusters with masses quite as large as the most massive in the Milky Way.  

The clusters that begin in low-mass sub-halos have an
average distance from the center of the galaxy of approximately 60 kpc, for those  within 150 kpc,
and the median radius is approximately 40 kpc. Therefore,
in contrast to the general distribution of stars formed at high redshift which are concentrated to the center
of the final dark matter halo, the stars formed in the lower mass sub-halos present at high redshift, and sufficiently massive
to be able to support star formation, the stars and clusters are distributed in a slightly
shallower radial distribution than the dark matter.

\begin{figure}
\begin{center}
\includegraphics[angle=-90,scale=0.7,trim=60 20 20 60, clip=true]{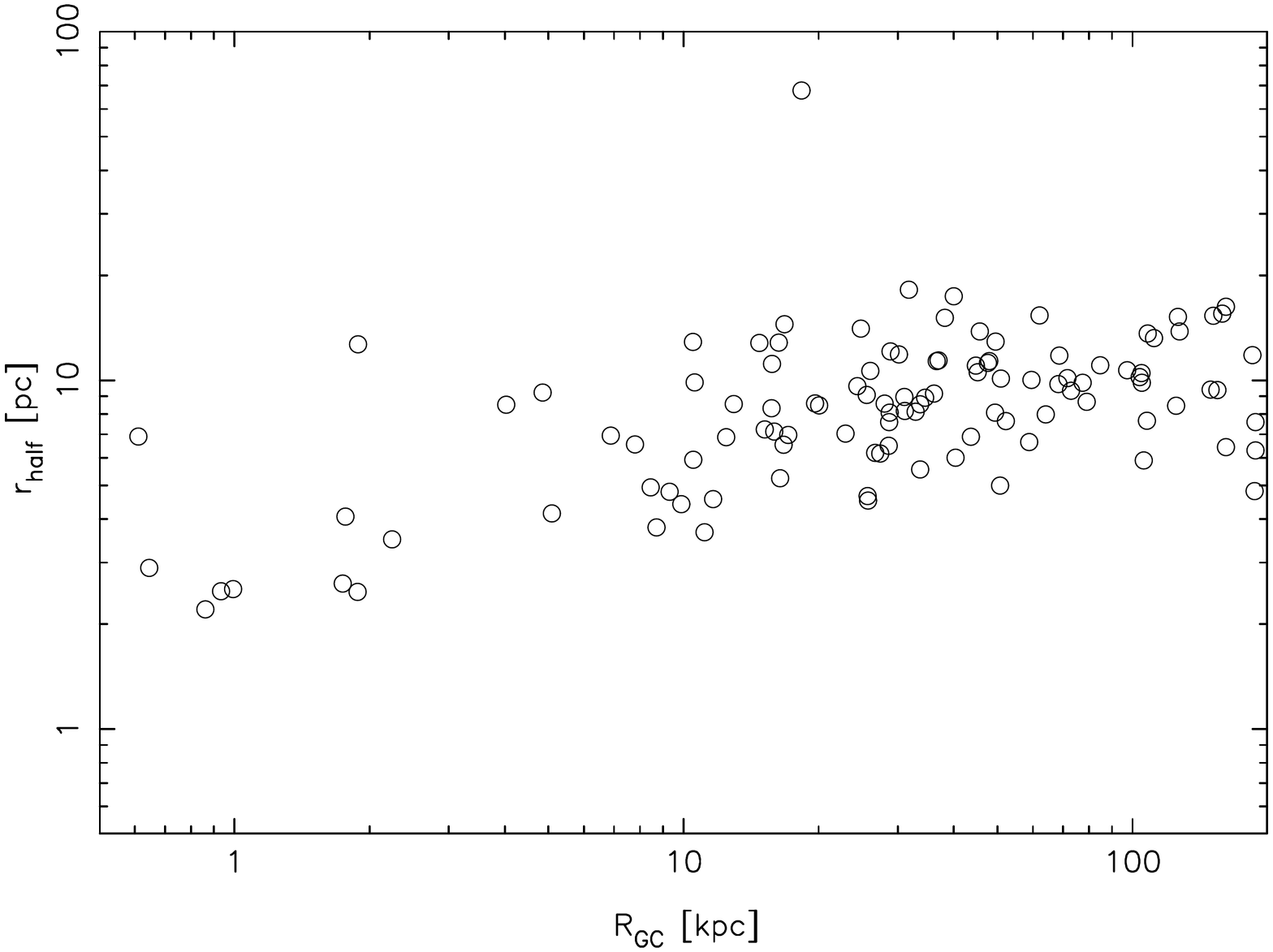}
\put(-110,-75){\includegraphics[angle=-90,scale=0.3,trim=60 20 20 60, clip=true]{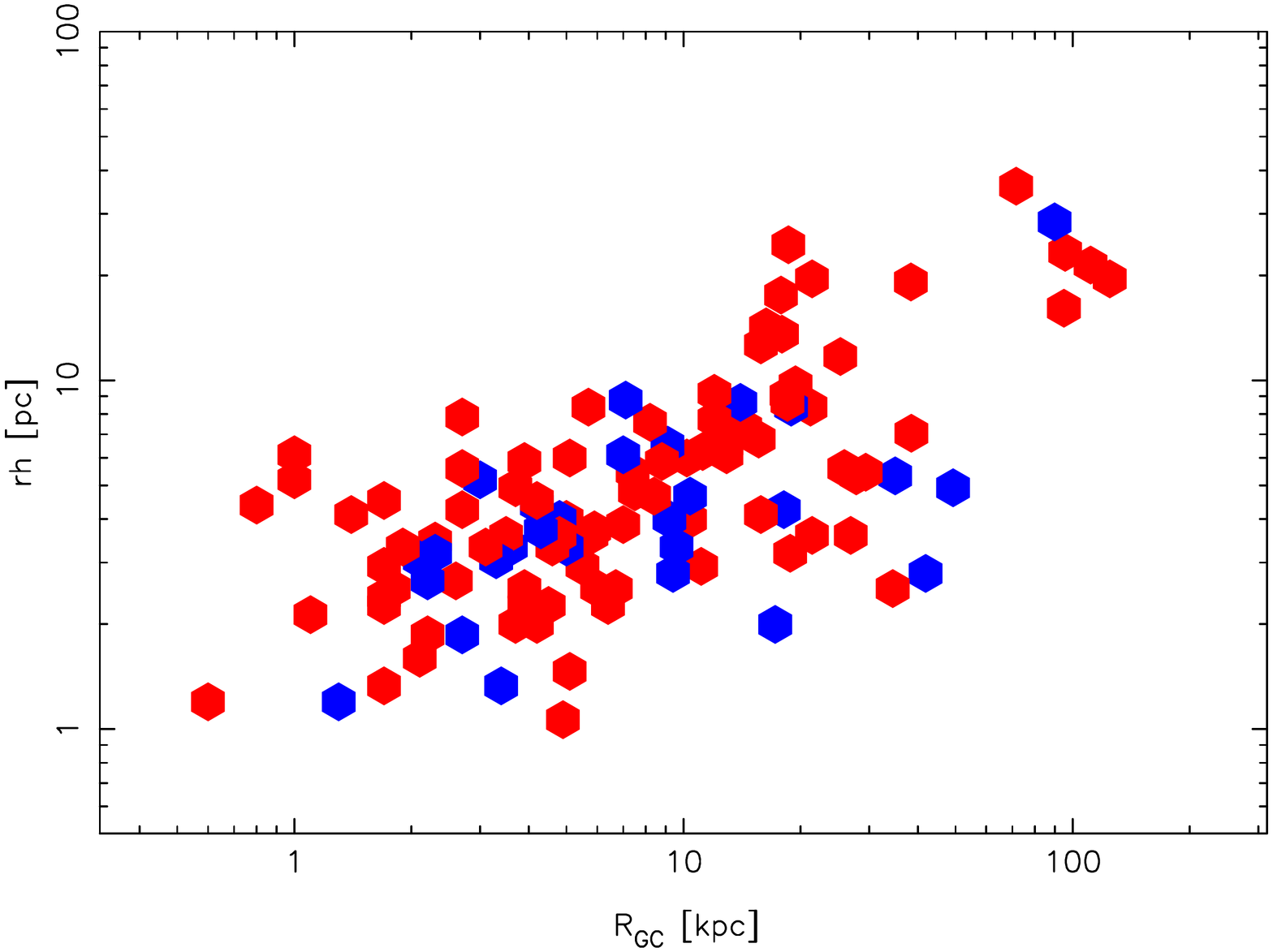}} 
\end{center}
\caption{
Cluster half mass radii as a function of their galacto-centric distance for the clusters at the
end of the redshift 8 simulation.
The Milky Way $\rm [Fe/H] \le -1$ clusters from the \citet{Harris:96} catalog are 
shown in the inset, with the colors indicating metallicity, as in Figure~\ref{fig_rm}.
}
\label{fig_rhR361}
\end{figure}

\begin{figure}
\begin{center}
\includegraphics[angle=-90,scale=0.7,trim=60 20 20 60, clip=true]{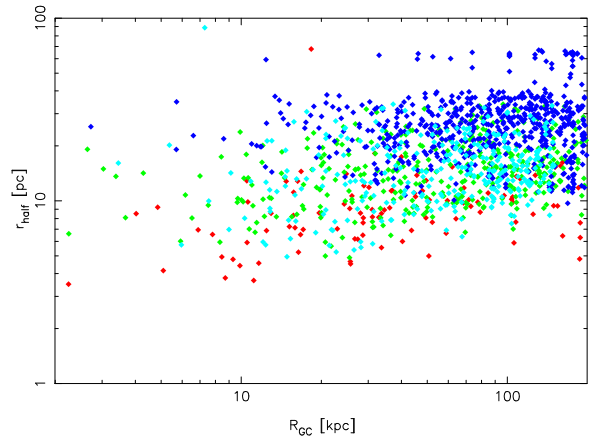} 
\end{center}
\caption{
Cluster half mass radii vs galactic radii for clusters
started at redshift 8 (red), 4.6 (green) and 3.2 (blue). 
The highest redshift clusters have the smallest half mass radius.
The redshift 3.2 clusters only have a small fraction with sizes
comparable to Milky Way halo clusters.
}
\label{fig_rrh}
\end{figure}

\begin{figure}
\begin{center}
\includegraphics[angle=-90,scale=0.7,trim=60 20 20 60, clip=true]{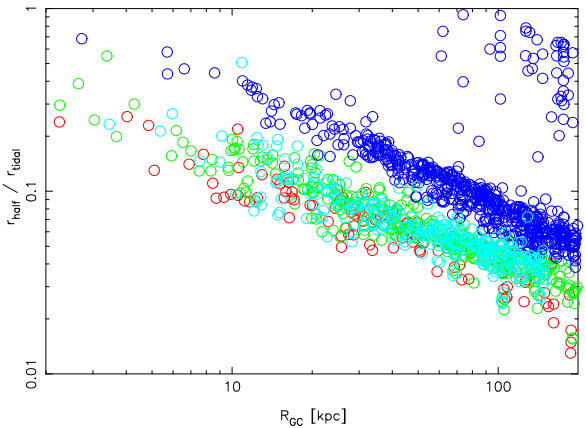} 
\end{center}
\caption{
The ratio of individual cluster half mass radius to its current location tidal radius as a function of galactic radius for clusters
started at redshift 8 (red), 4.6 (green) and 3.2 (blue). 
}
\label{fig_rrht}
\end{figure}

\begin{figure}
\begin{center}
\includegraphics[angle=-90,scale=0.7,trim=60 20 20 60, clip=true]{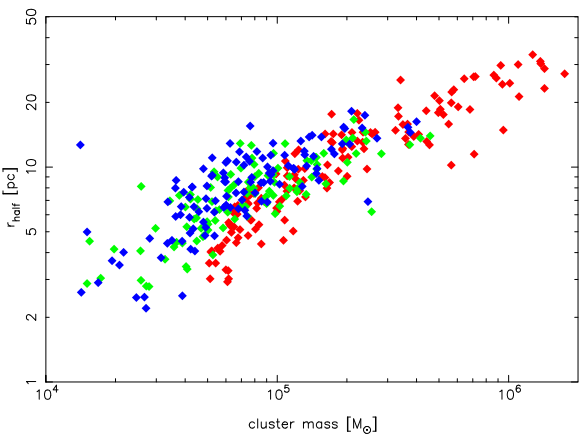}
\end{center}
\caption{
The half mass radius-mass relation for the clusters in the redshift 8 simulation at 
times  0.7 Gyr  (red), 7.5 Gyr (green) and 13.4 Gyr (blue). 
The relations varies little with time for clusters that remain bound, although the
clusters lose considerable mass with time.
}
\label{fig_mrh}
\end{figure}

\section{Cluster Half Mass Radius Dependence on Starting Redshift}

The simulated redshift 8 clusters 
substantially overlap in half-mass radii, Figure~\ref{fig_rhR361},
with those measured
in the Milky Way, shown as the inset in Figure~\ref{fig_rhR361}. 
The \citet{Harris:96} projected half-light radii have been multiplied by 1.33
to get to a 3D value \citep{Baumgardt:10} under the mass follows light assumption.  
After setting the $\rm [Fe/H] \geq -1$ Milky Way disk clusters aside, the distribution of very metal poor clusters, 
$\rm [Fe/H] \leq -2$
is broadly similar to the simulated clusters, although somewhat larger. The Milky Way contains a few clusters more massive than in
the simulation and are probably not adequately modeled in these simulations.

The tidal field sets an upper limit to the size of clusters. In principle gas cooling and star formation processes could produce significantly
smaller clusters, although as clusters become denser the internal star-star relaxation will increase. 
It is an interesting possibility that tidal fields, largely from the dark matter distribution in which they form,
 play the dominant role in determining the size of the clusters 
formed at high redshift which are later incorporated into the galactic halo.
Figures~\ref{fig_rrh} and \ref{fig_rrht} plot respectively, $r_h$, the half mass radius,  
and, $r_h/r_t$, the ratio of the half mass radius to the tidal 
radius, against the final time galactic radius for simulations started at redshift 8, 4.6 and 3.2.  
The redshift 4.6 simulation was run again starting with clusters 2.5 times smaller, but the clusters quickly
expand to fill their tidal radii in the sub-halo. Consequently
the results are not significantly dependent how  under-filling of  the tidal radius clusters are at the outset.
The half mass radius has a weak rise with galactic radius, $r_h\propto R_{GC}^{0.26\pm 0.2}$. 
The ratio $r_h/r_t$ in the final
galactic halo is a strong function of galactic radius, as a result of the dependence of tidal radius of galactic radius.
Most of the clusters have a half mass relaxation time
longer than an orbital time so the clusters do not change much in half mass size as they orbit, whereas the
local tidal radius varies around the orbit, leading to the strong dependence of the ratio on the galactic radius.
The half mass radii of the clusters at the beginning, halfway point, and end of the redshift 8 simulation
are shown in Figure~\ref{fig_mrh} which shows that the cluster half mass radii are largely determined
by the tidal fields present at the starting location of the clusters. Although clusters lose considerable
mass with time, Figure~\ref{fig_fmt}, much of it occurs at early times when the tidal fields are similar to those
at the start, Figure~\ref{fig_tides}. These simulations apply only to halo clusters where the strong tidal fields
of a baryonic galaxy can be ignored.

Figures~\ref{fig_rrh} and \ref{fig_rrht} also show that at any given galactic radius the clusters are ordered with 
decreasing $r_h$, and the ratio $r_h/r_t$, with increasing redshift at which they were started. 
For a more realistic simulation with a continuous cluster formation history the values
of $r_h$ would be smoothly distributed
and plausibly would resemble the same relation in the Milky Way \citep{Baumgardt:10}.
The width of the $r_h$ relation for a single
starting redshift is likely due to the range of tidal forces at the spread of the
initial radial locations of the clusters in their starting sub-halos, see Figure~\ref{fig_tides}.

It is likely that the starting redshift of formation and the initial 
orbital radius  at which the clusters are initially placed are somewhat degenerate quantities 
in determining final time $r_h$ values for individual clusters.
The redshift 3.2 starting conditions leads to a substantial fraction of the population have half mass radii
larger than are seen in the Milky Way globular cluster population. 
The simulation results here indicate that halo clusters must have formed
largely above redshift 4, for which there is some observational support \citep{KR:13}. 

Had the clusters been started closer to center of the initial sub-halos the
stronger tidal fields (assuming a cusped density profile) would lead to somewhat smaller tidal radii, hence, $r_h$ values. 
However, for a central $\rho(r)\propto r^{-1}$, $M(<r) \propto r^2$, the local tidal radius for a fixed
mass cluster varies slowly with radial location, $r_t\propto r^{1/3}$.
That is, a factor of two decrease in tidal radius requires a factor of 8 decrease in orbital radius 
in the sub-halo, which would move the clusters inside 100 pc. Even
at a 1 kpc starting radius the clusters will sink due to dynamical friction \citep{OLR:00,CDR:12}
which happens naturally in these simulations, as visible in a few clusters where the tidal
fields increase with time in Figure~\ref{fig_tides}. Since most sub-halos have only a single globular cluster at
the outset, friction drawing them to the center is not necessarily a problem, and sub-cluster merging
is working to place clusters in larger orbits in merged halos where dynamical friction is reduced.
Further simulations are required to clarify these issues.

The most massive star cluster allowed in the initial conditions was 
$2\times 10^6 \msun$ and the most massive clusters have a high relative 
mass loss rate than lower mass clusters, see Figure~\ref{fig_fmt}.
 It would be straightforward
to include more massive clusters in the simulation. Obtaining a cluster like M15/NGC7078
with a mass of nearly $10^6\msun$ and a half mass radius of 5 pc \citep{Baumgardt:10}, requires yet stronger
tidal fields or special formation conditions.

\section{Discussion}

The dynamical simulations here find that clusters formed at redshift 8 in higher mass dark matter halos are brought into the center 
of a Milky Way-like halo and either destroyed
at high redshift by the dark halo, or, later as the dense baryonic galaxy develops. 
On the other hand, the accreted lower mass dark halos put
their globular clusters into large radius orbits, 
an average galactocentric distance of about 60 kpc inside 150 kpc.  Most of the stars lost to early time tidal fields 
are eventually spread out over the halo, however late time
tidal streams are thin and are detectable at the current epoch. The Sylgr stream may
be an example \citep{IMM:19,RG:19}.

The relation between initial dark matter sub-halo mass and the orbits of clusters found at redshift 8 is not seen at redshift 4.6. 
The differing radial distribution is likely the
result of dynamical friction as sub-halos of varying mass fall into the main halo.  
The time scale for dynamical friction
is derived in \citet{BT:08}, Equation (8.13),
\begin{equation}
t_{df} = {1.17\over{\log{\Lambda}}} {M(r) \over {M_{sat}}} t_{cross},
\end{equation}
where $t_{cross}$ is the crossing time, defined as the travel time for one radian of orbit, $\log{\Lambda}$ is the Coulomb logarithm,
and $M(r)$ is the host halo mass inside radius $r$ (assumed to have $M(r)\propto r$), 
and $M_{sat}$  is the  sub-halo mass, approximated over an orbit as a constant mass satellite. 
The mass of the dominant halo declines drops quickly with increasing redshift.  In the
simulations here the most massive halo at redshift 8 is $0.94\times 10^{10}\msun$,
whereas at redshift 4.6 the most massive is $7.3\times 10^10\msun$, 7.8 times more massive.
In addition the dynamical time in the halo at high redshift is shorter because the universe is denser by a factor of $(1+z)^3$. 
Therefore, the same mass sub-halo falling into the dominant halo at redshift 8 has about a factor of 15 times more dynamical
friction per unit time than at redshift 5.6. 
At lower redshift dynamical friction time for the sub-halos becomes longer than the dynamical 
time for both high and low mass sub-halos so that all sub-halos merge into the main halo with relatively little loss
of orbital energy.

Observations show that lower luminosity galaxies host globular clusters that
tend to have lower metallicities than the field stars \citep{Lamers:17} with the Fornax dwarf
being a notable example \citep{LSB:12}.
Lower luminosity galaxies also have a higher number of globular clusters per unit galaxy luminosity \citep{GPG:10}.
Although the relation between galaxy luminosity and dark halo mass remains an active research
topic the strong correlation is well established in local group galaxies \citep{Brook:14} and much of the relation must have
been put in place at high redshift.  

Low metallicity stars and globular clusters appear
to have numbers as a function of metal abundance, Z, 
in quite good agreement with the simple one zone halo chemical evolution 
model. The model has a single parameter, the effective enrichment yield, $Y$, of a generation of stars,
and then predicts cumulative numbers 
$\rm N(<Z) \propto 1-\exp{(-Z/Y)}$  with a normalization that depends on when star formation ceases,
or, is matched to an observational data set 
\citep{Hartwick:76}. 
Depending on the details of calibrating to Milky Way halo numbers,  if globular clusters
trace the field star metallicity distribution, then 
3-10  clusters are expected below the 
apparent floor of $\rm [Fe/H]<-2.5$ if the clusters continue to be in proportion to the numbers 
of field stars with metallicity \citep{Youakim:20}.

The presence of unbound, chemically unique second generation globular cluster stars in the stellar halo 
of the Milky Way suggests that clusters as a population have lost 50+\% of their stellar mass over their lifetime \citep{Koch:19}.
In the tidally limited cluster framework advocated in this paper a large mass loss for halo clusters requires
that they were formed at greater than redshift 4, Figures~\ref{fig_fmt} and \ref{fig_fmt358},
when tides were sufficiently strong to drive substantial mass loss.

Some of the very rare extremely metal poor stars are in a thin distribution aligned with the Milky Way disk \citep{Sestito:19}.  
These stars could be 
the remnant of a very early gas disk of the Milky Way, or, the result of late infall of dwarf galaxy sufficiently massive
and dense to be pulled into the plane of the current disk \citep{TO:92,Walker:96,HC:97,VW:99}.  The distribution of extremely
metal poor stars will be affected by the buildup of the baryonic galaxy, which is another motivation 
to search for such stars at distances in the halo that are not significantly affected by the baryonic galaxy.

Observations at high redshift will eventually be able to measure the circular velocities of the dark matter halos
containing globular clusters. 
The circular velocities of the sub-halos are observational quantities
that are straightforward to compare to the same measurements in the simulations.
The low mass sub-halos have a mean peak circular velocity of $12-18$ \kms, 
comparable to the Fornax dwarf \citep{Walker:06}, although the clusters here are much denser. 
The high mass sub-halos have $26-55$ \kms. The clusters
are inserted about 20-30\% of this characteristic radius, so 
have a initial  velocity typically 30-50\% 
of the peak circular velocity of a halo.

\section{Conclusions}

The dynamical simulations of tidally limited star clusters reported here find that clusters that formed 
in the sub-halos present at redshift 8 have orbits in the resulting Milky Way-like halo that
depend on the  mass of the dark halo in which they formed.
The clusters that are formed in the more massive halos 
are deposited deep in the potential of the forming galactic halo, 
where the strong tides cause a lot of mass loss and leave the clusters at such small orbital radii 
that a baryonic bulge and disk would likely completely disperse the clusters.  
On the other hand, the clusters 
formed in the lower mass halos are spread out over the galaxy 
in a distribution which is similar or somewhat more extended than 
the dark matter density. The peak circular velocity of the lower mass halos
is $ 12-18$  \kms, 
The same starting procedure at redshift 4.6 leads to negligible differences between
the radial distribution of stars from the clusters in low and higher mass sub-halos. 

The sizes of the simulated clusters are limited by by their tidal radii, 
which are dependent on the tidal field present at the starting time
and location.
The clusters formed at higher redshifts will, on the average, be in stronger tidal fields and will
be the relatively smaller clusters.
The
simulations here find that tidally clusters need to have been formed at $z\ga 4$ to be a reasonable match to the Milky
Way distribution of sizes. Together these results indicate that a large fraction of the
halo clusters formed at $z>6$.

The simulations suggest that many of the first globular clusters to form, 
which will be the most metal poor, will be fairly evenly spread over
the entire galactic halos. Those that are dissolved could be indirectly found through 
the ongoing search for stellar streams.  As shown in the bottom left panel 
of Figure~\ref{fig_lh100}, stellar streams
from the high redshift clusters are visible in the inner galaxy and out to 100 kpc. 

\acknowledgements

This research was supported by  NSERC of Canada. Computations were performed on the niagara supercomputer 
at the SciNet HPC Consortium. SciNet is funded by: the Canada Foundation for Innovation; the Government of Ontario; 
Ontario Research Fund - Research Excellence; and the University of Toronto.

\end{document}